\documentclass[aps,prb,a4,reprint,twoside,floatfix,]{revtex4-1}   %[superscriptaddress]

\usepackage{graphicx}
\usepackage{setspace}
\usepackage{color,soul}
\usepackage{float}
\usepackage{amsmath}
\usepackage{amssymb}
\usepackage{amsfonts}
\usepackage{dcolumn}
\usepackage{epsfig}
\usepackage{subfigure}
\usepackage{bm}
\usepackage{setspace}   %Allows double spacing with the \doublespacing command
\usepackage{lipsum} % Add dummy text

\begin{document}

\title{Dynamics of photo-generated non-equilibrium electronic states in Ar$^+$ ion irradiated SrTiO$_3$}
\author{Dushyant Kumar, Z. Hossain and R. C. Budhani*}
\affiliation{Condensed Matter Low Dimensional Systems Laboratory, Department of Physics, Indian Institute of Technology, Kanpur 208016, India}
\email{rcb@iitk.ac.in}
\date{\today}

\begin{abstract}

 A metallic surface is realized on stoichiometric and insulating (100) SrTiO$_3$ by Ar$^+$ - ion irradiation. The sheet carrier density and Hall mobility of the layer are $\sim$\,$4.0$ $\times$ $10^{14}$ $/cm^2$ and $\sim$\,$2$ $\times$ $10^3$ $cm^2/Vs$ respectively at 15 K for the irradiation dose of $\sim$\,4.2 $\times \:\:10^{18}$ $ions/cm^2$. These samples display ultraviolet light sensitive photoconductivity (PC) which is enhanced abruptly below the temperature ($\approx$\,100 K) where SrTiO$_3$ crystal undergoes an antiferrodistortive cubic-to-tetragonal ($O_h^1$ $\rightarrow$ $D_{4h}^{18}$) structural phase transition. This behaviour of PC maps well with the temperature dependence of dielectric function and electric field induced conductivity. The longevity of the PC-state also shows a distinct change below $\approx$\,100 K. At $T > 100$ K its decay is thermally activated with energy barrier of $\approx$\,36 meV, whereas at $T < 100$ K it becomes independent of temperature. We have examined the effect of electrostatic gating on the lifetime of the PC state. One non-trivial result is the ambient temperature quenching of the photo-conducting state by the negative gate field. This observation opens avenues for designing a solid state photo-electric switch. The origin and lifetime of the PC-state are understood in the light of field effect induced band bending, defect dynamics and thermal relaxation processes.

\end{abstract}
\maketitle

\section{Introduction}

Understanding the physics and chemistry of complex oxides and their interfaces has been a topic of extensive study over the last few decades because of their novel properties many of which are technologically very important.\cite{mannhart2010oxide} Among many oxides SrTiO$_3$ (STO) is one of the most widely investigated, particularly from the perspective of oxide-based electronics.\cite{zhao2006ultraviolet,kan2006blue} Strontium titanate is a wide band gap semiconductor with an indirect gap of 3.25 eV,\cite{van2001bulk} whereas its direct band gap is 3.75 eV.\cite{zhang2013unusual} Some fascinating characteristics of STO include high electron mobility ($\sim$\,$10^4$ $cm^2/Vs$) at liquid helium temperature on electron doping,\cite{Tufte1967electronmobility,lee1971electronic}
%with many other interesting properties like insulator-metal transition around the critical carrier density of $\sim10^{18}\:\:cm^{-3}\:\:$,\cite{Lee}
superconductivity at low temperatures in the doped state,\cite{ueno2008electric} electronic tunability of surface states by electrostatic gating,\cite{lee2011electrolyte} blue light emission on Ar$^+$ - ion irradiation\cite{kan2006blue} and the observation of a two-dimensional electron gas showing magnetic and superconducting correlations at its interface with some other non-conducting perovskite oxides.\cite{lee2011electrolyte,Tufte1967electronmobility,ohtomo2004high,herranz2007high,kalabukhov2007effect, siemons2007origin,hwang2012emergent,reyren2007superconducting,brinkman2007magnetic,biscaras2012two} The many unique properties of STO have a strong bearing on its dielectric function which increases rapidly below $\simeq 120 K$ and also becomes electric field dependent.\cite{ChristenPhysRevB,neville1972permittivity} The cubic perovskite structure of STO undergoes antiferrodistortive (AFD) cubic-to-tetragonal ($O_h^1$ $\rightarrow$ $D_{4h}^{18}$) phase transition (PT) on cooling below 105 K.

Many groups have studied the photoconducting state in nominally pure single crystals of SrTiO$_3$\cite{katsu2000anomalous,zhang2013unusual,rossella2007photoconductivity,Jin2013jap} with a particular attention to the effect of AFD-PT on photoconductivity. However, a clear connection between the two has not emerged as the very high resistance of single crystals of STO makes it difficult to examine thoroughly the dark conductivity and photoconductivity under different probing fields.

Recently, Rastogi et al., have studied the photoconductivity (PC) in LaAlO$_3$/SrTiO$_3$ and LaTiO$_3$/SrTiO$_3$ based heterostructures,\cite{rastogi2010electrically,rastogi2012photoconducting,rastogi2012solar,rastogi2014enhanced} where they found persistent photoconductivity (PPC) and suggested a role of oxygen vacancies,\cite{rastogi2012solar,rastogi2012photoconducting} created in STO during the film deposition, in the stability of the PPC. These exciting results have increased interest in the studies of oxygen deficient STO even more. Oxygen vacancies can be created in STO by high temperature annealing in vacuum\cite{muller2004atomic,yamada1973point}, using hydrogen plasma cleaning method\cite{Takahashi2003H-plasma} and/or through Ar$^+$ - ion irradiation.\cite{kan2006blue,bruno2011anisotropic} In these processes, the excess electrons associated with the oxygen vacancies reduce the oxidation state of some of the Ti atoms from +\,4 to +\,3 making the material conducting.

In this report, we examine the PC behavior of SrTiO$_3$ single crystal slabs whose surface has been made conducting by ion irradiation. These studies are augmented by examination of the effects of electrostatics gating on the post-illumination recovery process. We find a strong signature of AFD-structural phase transition in the photo-induced conductivity as it undergoes an abrupt enhancement below $\approx$\,100 K. At room temperature, the negative gate field accelerates the recovery process of the PC-state leading a complete relaxation to its ground state. However, no such acceleration was observed at 100 K. The former case offers the opportunity to create a visible blind near UV electrostatic gate tunable photo-detector.

\section{experimental details}

\begin{figure}[h]
\begin{center}
\includegraphics [ trim=0cm 0cm 0cm 0cm, width=8.6cm, angle=0 ]{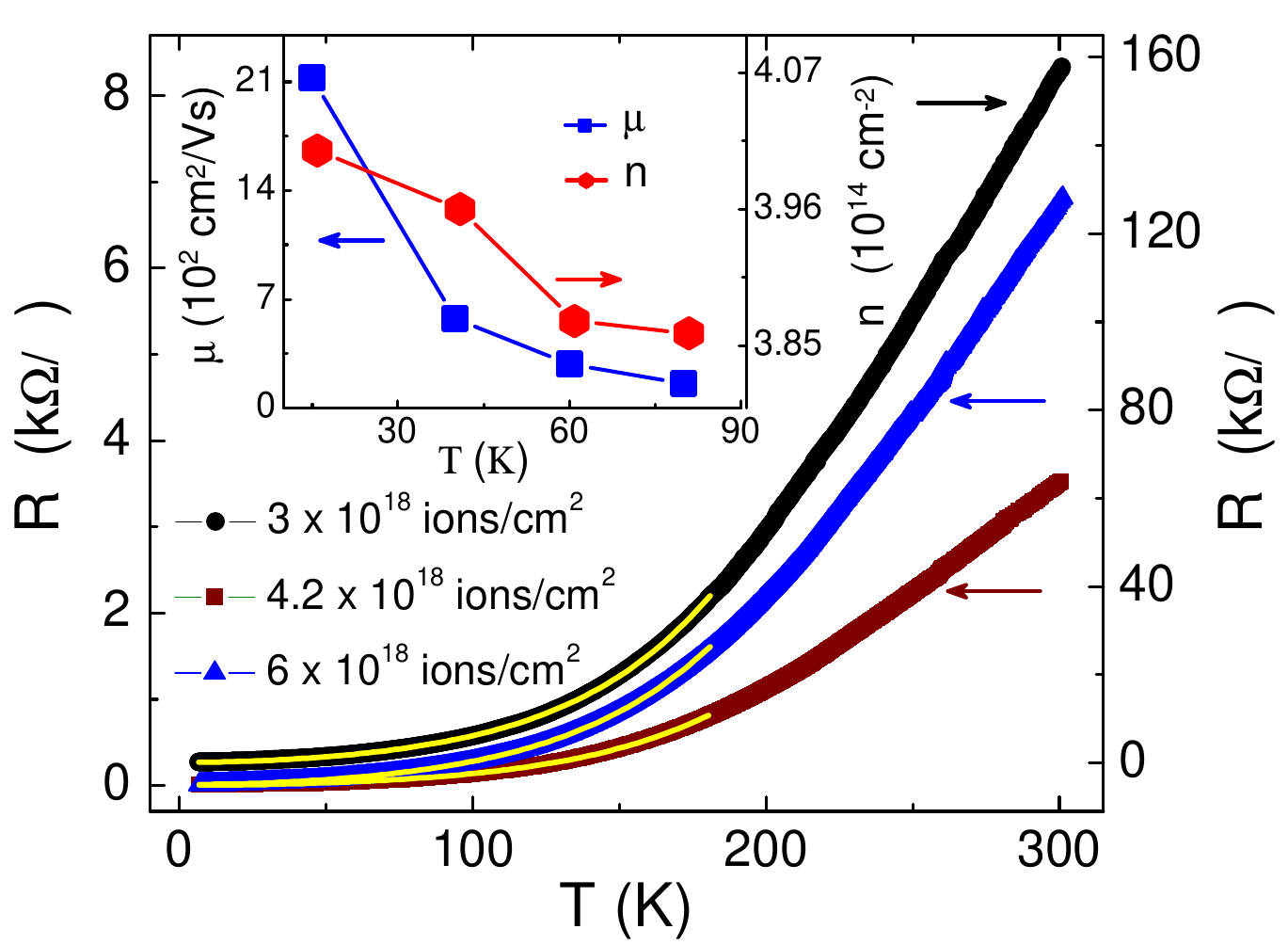}%
\end{center}
\caption{\label{fig 1} (Color online) Temperature dependence of the sheet resistance $R_{\Box}$ of three STO (100) slabs irradiated at different dosage of Ar$^+$ ions. The solid red lines are best fits to the expression $R_{\Box} = R_0 + R_1T^2 + R_2T^5$, within the temperature range of 10 K to 180 K. The inset shows mobility ($\mu$) and sheet carrier density ($n_\Box$) of the sample irradiated at $\sim$\,$4.2 \times 10^{18}$ $ions/cm^2$ as a function of temperature in the range 15 K to 80 K.}
\end{figure}

The SrTiO$_3$ (100) substrates acquired from Crystal GmbH Germany had no signatures of any measurable photoconductivity on illumination with a quartz halogen lamp. The Ar$^+$ - ion irradiation experiments on 0.5 mm thick and optically polished STO slabs were carried out in a Kauffman type ion source operated at $\sim$\,$8.5\: \times\: 10^{-4}$ mbar Ar-pressure. The typical acceleration voltage and ion current used for irradiation are 550 V and 1.5 $mA/cm^2$ respectively. The Ar$^+$ - ion irradiation provides a direct method to fabricate electron gases within a thickness of nanometers on the surface of STO, which can be estimated on the basis of ion penetration depth (L) using the empirical formula,\cite{schultz2007relaxation}\\
L = 1.1 $\times$ $\frac{E^{2/3}W} {\rho(Z_i^{1/4} \:+\: Z_t^{1/4})^2}$, \\
where, W (in amu), $Z_t$ and $\rho$ are the atomic weight, atomic number and density of the target respectively; E is the ion energy in eV and $Z_i$ is the atomic number of the ions. The penetration depth L of ions in our case comes out to be $\sim$\,$35 \:\dot{A}$. Therefore, we expect the conducting layer thickness to be of this order. The STO (100) substrates are irradiated in three different patterns; $3 \times 3$ $mm^2$, $1.5 \times 5$ $mm^2$ and 100 $\mu m$ $\times$ 5 mm depending on the requirements of a specific measurement. Electrical transport measurements (mobility ($\mu$) and sheet carrier density ($n_{\Box}$)) were performed in the Van der Pauw configuration, where Ag/Cr electrodes were deposited on the corners of the $3 \times 3$ $mm^2$ irradiated STO using shadow mask, whereas the sheet resistance was measured in the standard four-probe linear geometry as well as in the Van der Pauw configuration. For the PC measurements, we have used $1.5 \times 5$ $mm^2$ as well as 100 $\mu m$ $\times$ 5 mm irradiated channels by depositing Ag/Cr contact pads through a shadow mask, leaving a gap of $\approx$\,2 mm between the contact pads. Both samples showed similar PC behavior. The carrier density in irradiated channels has been tuned electrostatically by a metallic gate deposited on the back side of the STO substrate. A typical field-effect-transistor (FET) geometry of drain, source and gate electrodes used for the PC and gate field effect measurements is shown in the inset of Fig. 2. The ultraviolet (UV) light intensity of $\sim$\,$30$ $\mu W/cm^2$ from an unfiltered quartz halogen lamp was used to illuminate the samples. The spectral profile of the lamp reported earlier\cite{Rastogi2013thesis} consist of UV range (350 nm - 400 nm), which is about 3.5$\%$ of the total integrated intensity. All temperature dependence measurements were performed in vacuum ($<$\,2\,$\times$\,10$^{-2}$ mbar) in a close cycle helium cryostat having a quartz window for optical access with $\sim$\,90\% transmission in the spectral range of 250 nm to 850 nm. The samples degrade with time if left in air for few days. The results are reproducible if the samples are kept in vacuum and preferably at low temperatures.

\section{Results}
\subsection{Electrical transport}

Figure 1 shows the sheet resistance, $R_{\Box}$, measured in a four-probe configuration of the Ar$^+$ - ion irradiated STO at the cumulative doses of $\sim$\,3 $\times \:\:10^{18}$, 4.2 $\times \:\:10^{18}$ and 6 $\times \:\:10^{18}$ $ions/cm^2$. The temperature dependent $R_{\Box}$ measured from room temperature down to 10 K shows metallic response with a quadratic T-dependence at low temperatures. The sheet resistance at 300 K decreases from 157 $k\Omega$ to 3.5 $k\Omega$ on increasing the dose from $\sim$\,3 $\times \:\:10^{18}$ $ions/cm^2$ to 4.2 $\times \:\:10^{18}$ $ions/cm^2$, which can be attributed to the enhanced oxygen vacancies. However, a further increase in the dose to 6 $\times \:\:10^{18}$ $ions/cm^2$ leads to doubling of the sheet resistance from 3.5 $k\Omega$ to 7 $k\Omega$, which may be due to excessive vacancy creation and possibly amorphization of the surface. We have therefore used the intermediate dose of $\sim$\,4.2 $\times \:\:10^{18}$ $ions/cm^2$ to irradiate the samples for all other measurements. The $R_{\Box}(T)$ curves shown in Fig. 1 have been fitted to the expression $R = R_0 + R_1T^2 + R_2T^5$, over the temperature range of 10 to 180 K. Here, $R_0$ is the residual resistance, $R_1$ represents the contribution from $e^-$ - $e^-$ interactions while $R_2$ originates from electron - phonon scattering.\cite{Jin2013jap,Zhao2000PRL} The solid red lines shown in the figure are the fit to the data, which yield the parameters $R_1$ $\sim$\,$0.5$, 0.01 and 0.02 $\Omega /K^2$ whereas $R_2$ $\sim$\,$1$ $\times \:\:10^{-7}$, 2 $\times \:\:10^{-9}$ and 4 $\times \:\:10^{-9}$ $\Omega /K^5$ for the cumulative ions doses of $\sim$\,3 $\times \:\:10^{18}$, 4.2 $\times \:\:10^{18}$ and 6 $\times \:\:10^{18}$ $ions/cm^2$, respectively. The dominant $T^2$ dependence of $R_{\Box}$ with a weak contribution from electron-phonon scattering is evident for the metallic nature of reduced STO. The LaTiO$_3$/SrTiO$_3$ interface has been reported to exhibit similar quadratic temperature dependence of $R_{\Box}$.\cite{rastogi2012photoconducting} The temperature dependence of $\mu$ and $n_\Box$ measured in Van der Pauw geometry over the range of 15 to 80 K is shown in the inset of Fig. 1. We note that the $n_\Box$ $\sim$\,$4.0$ $\times$ $10^{14}$ $/cm^2$ while $\mu$ comes out to be $\sim$\,$21$ $\times$ $10^2$ $cm^2/Vs$ at 15 K, which are in good agreement with the reported values at low temperatures for oxygen deficient strontium titanate.\cite{lee1971electronic,Ngai2010PRB} The lower mobility seen here compared to the mobility of LaAlO$_3$/SrTiO$_3$ interface ($\sim$\,$10^4$ $cm^2/Vs$)\cite{herranz2007high} is presumably due to abundance of defects in the irradiated STO and also due to a diffuse interface between the conducting and non-conducting part of STO as seen in the high resolution microscopy.\cite{sanchez2013characterization}

%\begin{figure}[h]
%\begin{center}
%\includegraphics [ trim=0cm 0cm 0cm 0cm, width=8cm, angle=0 ]{figure2.eps}%
%\end{center}
%\caption{\label{fig 2} (Color online) Cross-sectional transmission electron microscopy images of (A) Ar$^+$ - irradiated (36 W for 5 min) single crystal of STO (001) taken from Santolino et al.\cite{sanchez2013characterization} and (B) 15-uc-thick LaAlO$_3$ film grown on SrTi$_3$ by Reyren et al.\cite{reyren2007superconducting}. A damaged surface is clearly observed on Ar$^+$ - irradiated STO whereas the LAO/STO interface shows the coherent growth.}
%\end{figure}

\begin{figure}[h]
\begin{center}
\includegraphics [ trim=0cm 0cm 0cm 0cm, width=8cm, angle=0 ]{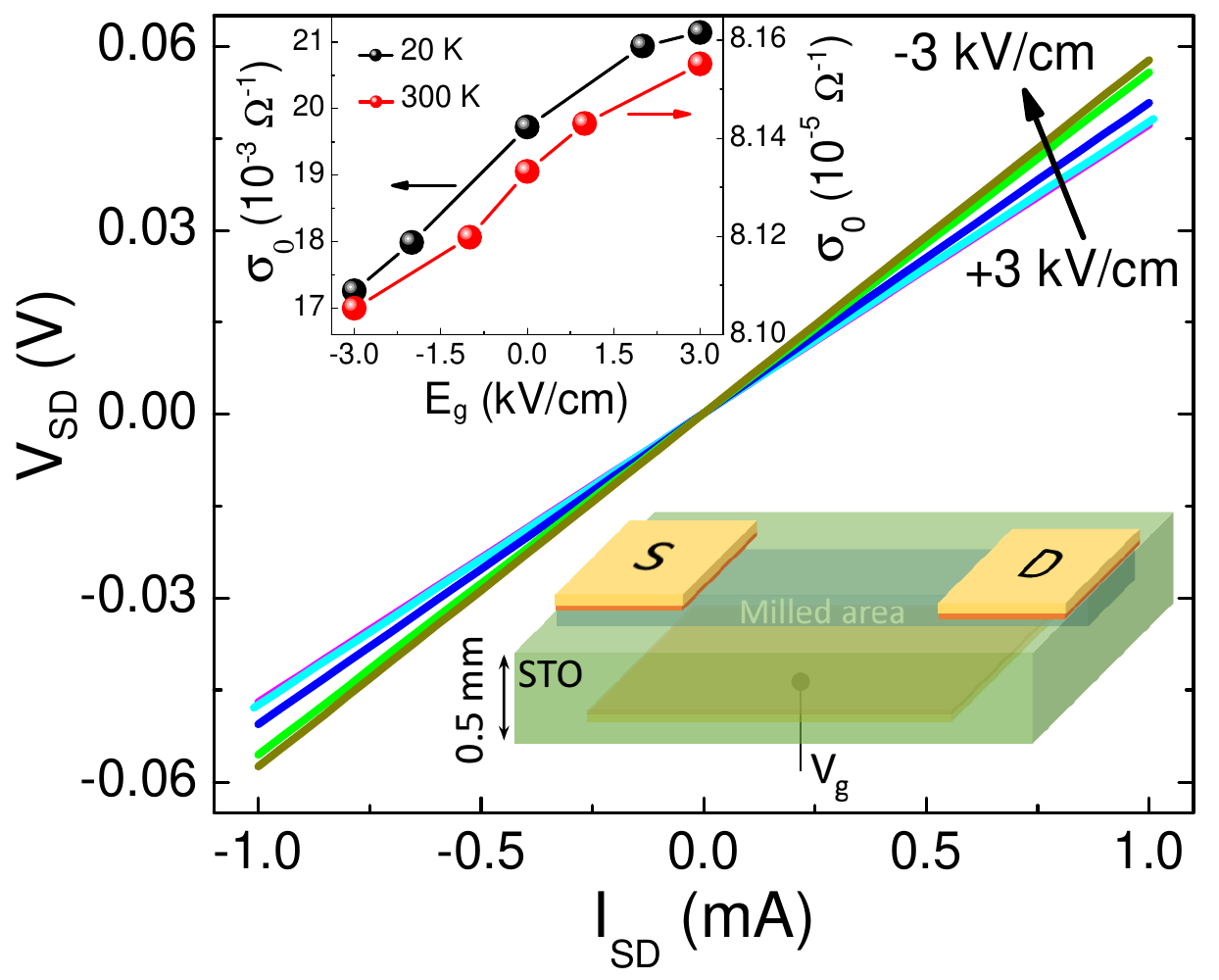}%
\end{center}
\caption{\label{fig 2} (Color online) Source-drain voltage ($V_{SD}$) as a function of current ($I_{SD}$) at different gate fields varying from -\,3 $kV/cm$ to +\,3 $kV/cm$. The data shown in the figure are taken at 20 K for the sample of ion dose $\sim$\,$4.2 \times 10^{18}$ $ions/cm^2$. Sketch of a typical FET-structure is shown in the lower inset. The electrodes are made of Ag/Cr bilayers deposited through a shadow mask. The upper inset shows zero bias conductance at 20 K and 300 K calculated from the slope of the $I_{SD}$-$V_{SD}$ curves at the origin.}
\end{figure}

\subsection{Gate effect on dark conductivity}

Several source-drain current-voltage (I-V) curves of the FET structures were collected for positive and negative gate fields at various temperatures between 20 and 300 K. The conductance of the channel has a strong dependence on the polarity of the bias field similar to as seen in LTO/STO and LAO/STO interfaces\cite{biscaras2014limit}. The I-V curves at 20 K under different gate fields between 0 and $\pm$\,3 kV/cm are shown in Fig. 2. The curves are linear throughout the temperature range emphasizing that the contacts are ohmic and the channel is metallic. We have measured the zero bias conductance, $\sigma_0$ = $(dI/dV)_{I_{SD}=0}$, from such linear curves for all gate fields at different temperatures from 20 to 300 K. For comparison, the 20 K and 300 K data of $\sigma_0$ have been plotted in the upper inset of Fig. 2. It shows enhancement by a factor of $\sim$\,$200$ upon lowering the temperature from 300 K to 20 K. The behaviour of the source-drain voltage, $V_{SD}$, as a function of gate field $E_G$ varying from 0 to $\pm$\,3 kV/cm is summarized in Fig. 3. For accuracy, every complete cycle of $E_g$ going from 0 to +\,3 kV/cm to -\,3 kV/cm and then back to 0 has been repeated three to four times at a fixed T-value. The normalized resistance $R/R_0$ vs $E_g$ plots are shown in the inset (a) of Fig. 3, where $R_0$ is the resistance at $E_g$ = 0. The curves are linear in both directions of $E_g$ at all temperatures and no hysteresis has been observed. The increase in resistance with varying $E_g$ from +\,3 kV/cm to -\,3 kV/cm indicates an n-type behavior of reduced STO, where the negative gate field reduces the charge carriers leading to increase in resistance. We have calculated the percentage change in the channel resistance defined as,\\
$\mid\Delta R_{E_g}\mid\:\: = \frac{\mid R(E_g\: =\:\: 0) \:-\: R(E_g\: =\:\:\pm\,3)\mid} {R(E_g\: =\:\: 0)} \times 100$,\\
where, $R(E_g = 0)$ and R($E_g = \pm\,3$) are the channel resistances at zero and $\pm$\,3 kV/cm gate fields, respectively. Although, the channel resistance decreases (increases) as an effect of positive (negative) gate field, the difference between magnitudes of the change in channel resistance by +\,3 kV/cm and -\,3 kV/cm is below 10\% throughout the T-range. The main panel of Fig. 3 shows the temperature dependence of the percentage $\Delta R_{E_g}$. A large enhancement in $\Delta R_{E_g}$ below T $\approx$\,100 K is evident in the figure. In Fig. 3 we also show the temperature dependence of the dielectric function ($\epsilon$) of undoped SrTiO$_3$ taken from Ang et al.\cite{ang2000effect} One can clearly see a parallel in the growth of $\epsilon$ and $\Delta R_{E_g}$ on lowering the temperature.
\begin{figure}[h]
\begin{center}
\includegraphics [ trim=0cm 0cm 0cm 0cm, width=8.6cm, angle=0 ]{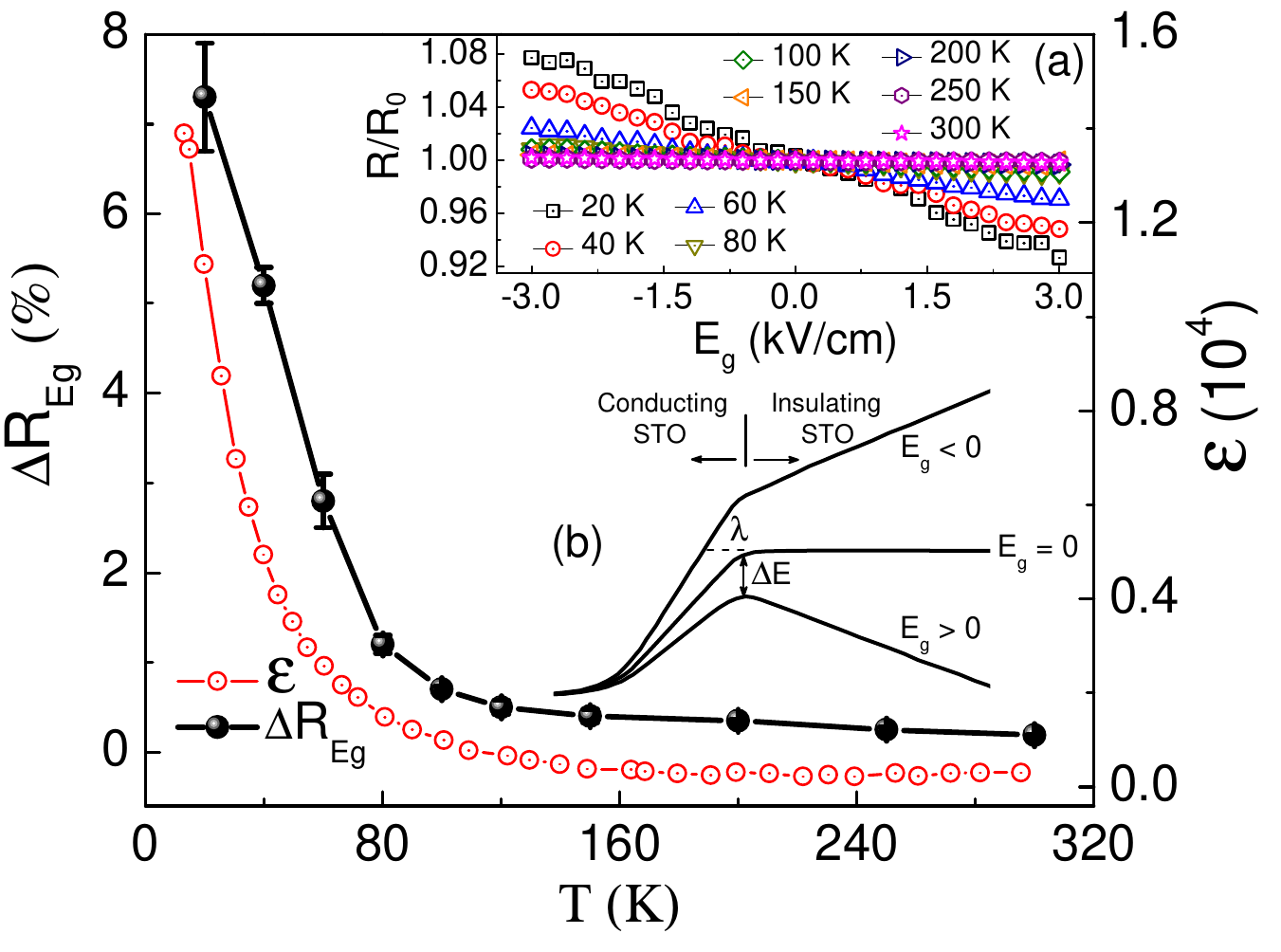}%
\end{center}
\caption{\label{fig 3} (Color online) Temperature dependence of the percentage change in resistance of sample with ion dose of $\sim$\,$4.2 \times 10^{18}$ $ions/cm^2$; $\Delta R_{E_g} (\%) = \:\mid R(E_g = 0) - R(E_g = \pm\,3) \mid/ {R(E_g = 0)}$ $\times$ 100, due to gate field is shown by black spheres with error bars in the main panel of figure. The red open circles show Ang et al.\cite{ang2000effect} data for the temperature dependence of dielectric constant ($\epsilon$) of undoped SrTiO$_3$. The $\epsilon$ and $\Delta R_{E_g}$ grow in parallel on lowering the temperature. Although, the positive and negative gate fields change $R_{\Box}$ in opposite manner, the magnitude of $\Delta R_{E_g}$ in two cases differs within 10\%. Inset (a) shows the normalized resistance as a function of gate field measured at several temperatures between 300 K and 20 K. An illustration of band bending at the interface due to gate field is sketched in the inset (b).}
\end{figure}
\subsection{Photo-induced changes in sheet resistance}

The sheet resistance of reduced STO during and after photo-illumination with a broad band source of net UV flux on the sample of $\sim$\,$30$ $\mu W/cm^2$ was measured at several temperatures between 15 K to 300 K. The typical acquisition time of $R_{\Box}$(t) profiles was $\sim$\,$2 \times 10^3$ s after turning off the illumination, while data for longer times were also recorded occasionally. We now define the normalized relative change in resistance upon photo-exposure as $\Delta R_N(t)$ = [$R_0$\:\:-\:\:R(t)]/$R_0$, where $R_0$ is the resistance at t = 0 (resistance in dark) and R(t) is resistance at time t. The temporal dependence of $\Delta R_N(t)$ has been shown in the inset of Fig. 4. The samples were exposed to UV light for 5 minutes starting from point A to point B after which the isothermal recovery of $R_{\Box}$(t) is monitored with t. This temporal evolution of $\Delta R_N(t)$ under illumination has two components; a rapid fall, with timescale of the order of milliseconds, followed by a gradual decrease. The post illumination recovery process also shows similar behavior with a sudden jump in R-value followed by a much slower tail.
\begin{figure}[h]
\begin{center}
\includegraphics [ trim=0cm 0cm 0cm 0cm, width=8cm, angle=0 ]{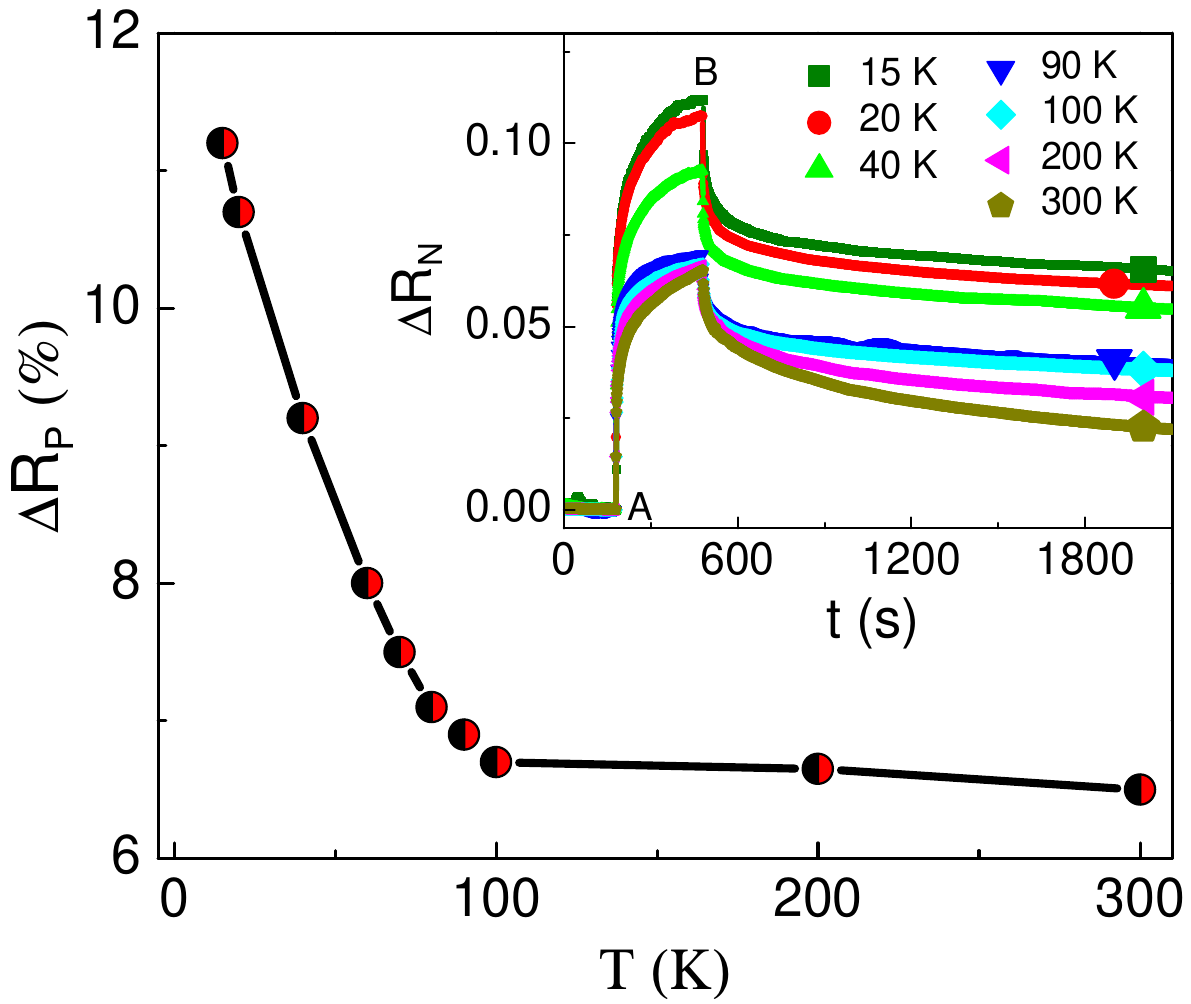}%
\end{center}
\caption{\label{fig 4} (Color online) The photo-induced percentage change in resistance as a function of temperature; $\Delta R_P (\%) = [R_0 - R(t = t_1)]/R_0$ $\times$ 100. Here $t_1$ is the time when UV light is turned off and $R_0$ is the resistance in dark (at t = 0). The photo-induced conductivity shows an abrupt enhancement below $\sim$\,100 K. Inset shows the temporal dependence of the normalized relative change in resistance under and after photo-illumination, $\Delta R_N(t)$ = [$R_0$-R(t)]/$R_0$, where $R_0$ is the resistance at t = 0 and R(t) is resistance at time t, realized at various temperatures.}
\end{figure}

\begin{figure}[h]
\begin{center}
\includegraphics [ trim=0cm 0cm 0cm 0cm, width=8cm, angle=0 ]{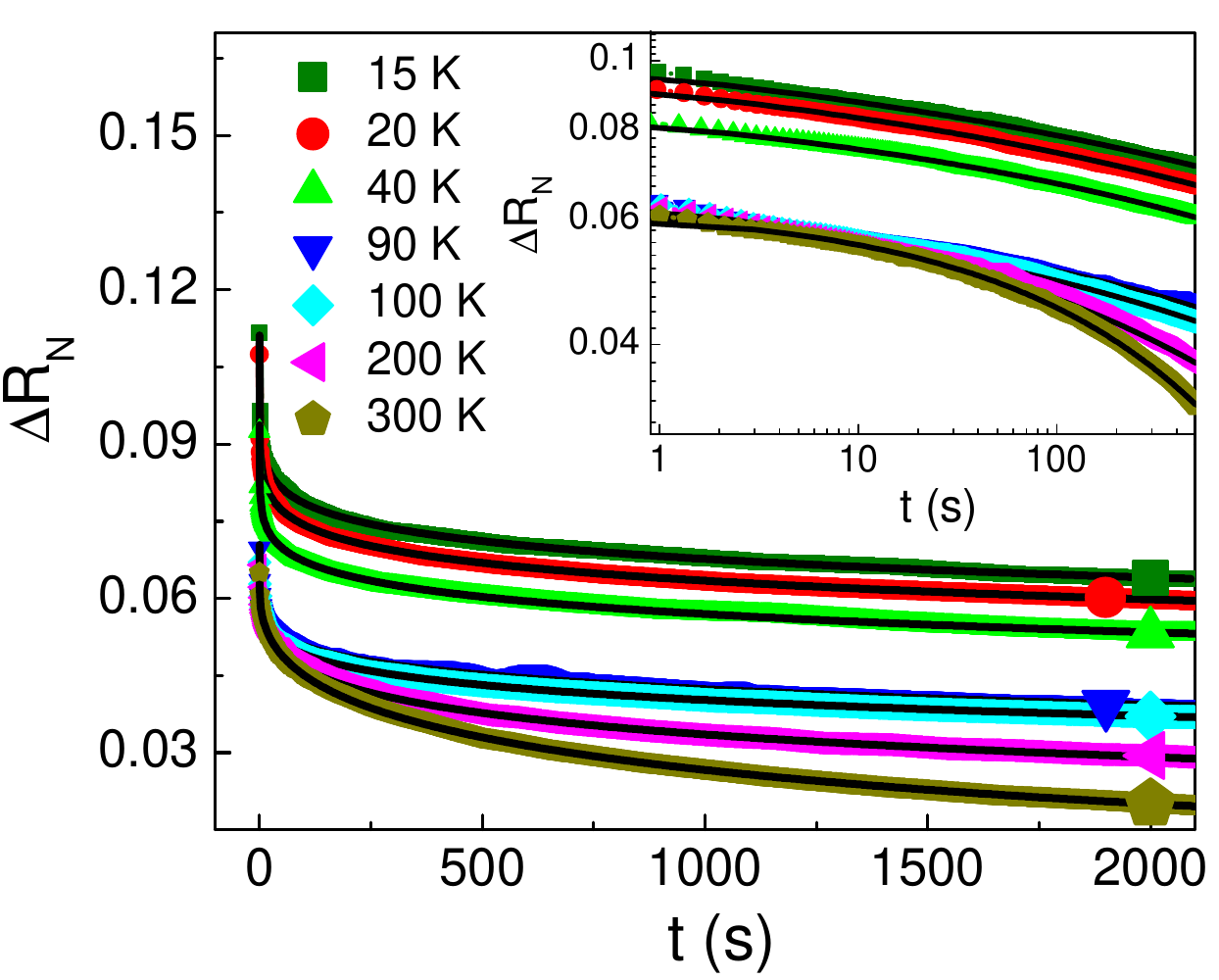}%
\end{center}
\caption{\label{fig 5} (Color online) Time evolution of $\Delta R_N(t)$ = [$R_0$-R(t)]/$R_0$, after switching off photo-exposure. The recovery curves show large persistent photoconductivity. The solid black lines are the best fits to the stretched exponential $\Delta R_N(t)$ $\sim$ $\exp[-(\frac{t}{\tau})^\beta]$. Inset shows the $\Delta R_N(t)$ profiles with fittings below t = 500 sec on a log-log scale.}
\end{figure}

\begin{figure}[h]
\begin{center}
\includegraphics [ trim=0cm 0cm 0cm 0cm, width=8cm, angle=0 ]{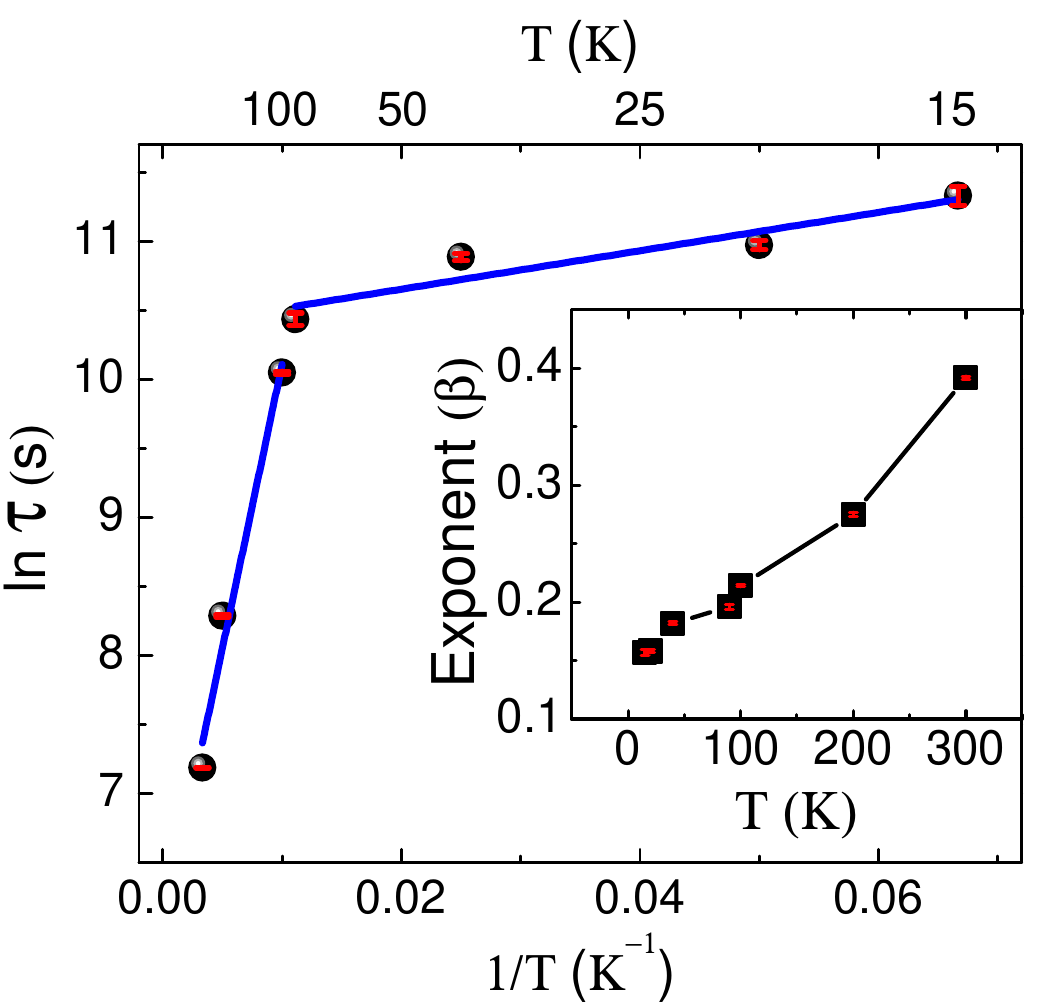}%
\end{center}
\caption{\label{fig 6} (Color online) ln ($\tau$) as a function of 1/T with error bars is plotted in the main graph whereas the temperature dependence of the decay exponent, $\beta$ and the error bars are shown in the inset. The T-dependence of the decay time constant $\tau$, is fitted to Arrhenius equation with two different activation energies well separated in temperature at $\sim$\,100 K. The fits are shown as solid blue lines.}
\end{figure}
For a better understanding of the photo-response, the net percentage change in resistance under photo-illumination was defined as $\Delta R_P$ = $\Delta R_N(t = t_1)$ = [$R_0 - R(t = t_1)]/R_0$ $\times$ 100, where $t_1$ is the time when UV light is turned off and $R_0$ is the resistance in dark. The $\Delta R_P$ has been plotted as a function of T in the main graph of Fig. 4. The curve shows two very distinct regions which are well separated around 100 K. There is a gradual increase in $\Delta R_P$ on cooling from 300 K to 100 K, while below 100 K the $\Delta R_P$ increase sharply. The $\Delta R_P$($\%$) at 15 K is nearly twice the value at 300 K. This dramatic change in the shape of $\Delta R_P$ vs T curve strongly suggests that there are two different mechanism responsible for the different PC behavior seen below and above 100 K.

To further analyse the photo response, we monitor the post illumination recovery of the system. Figure 5 shows the behaviour of $\Delta R_N(t)$ at several temperatures measured over a wide time span. As full recovery implies $\Delta R_N(t)$ $\rightarrow$ 0, it is clear from Fig. 5 that the recovery process slows down significantly at lower temperatures and becomes independent of T below $\approx$\,100 K. The decay profiles can be fitted to the stretched exponential given by Kohlrausch.\cite{Kohlrausch}\\
$\Delta R_N(t)$ $\sim$ $\exp[-(\frac{t}{\tau})^\beta]$ \\
Here, $\tau$ is the characteristic decay time constant of the process, while $\beta$ the decay exponent takes a value $\leq$ 1. Here the equality stands for the Debye relaxation processes which are governed by a single activation energy. The stretched exponential fits to the data are shown as solid black lines in Fig. 5. The temperature dependence of the parameters $\tau$ and $\beta$ is shown in Fig. 6. For the relaxation time, one can notice two distinct temperature dependent regions; one at low T (below $\sim$\,100 K) and other at high T (above $\ 100$ K) indicating two activation energies, which can be expressed by the Arrhenius form,\\
$\tau \sim \tau_0\exp[-\frac{E_T} {K_BT}]$,\cite{nelson1977long,mooney1990deep,Lin1990PRB}\\
where, the activation energy $E_T$ is for the capture of electrons at the defect centers measured from the quasi-Fermi-level in the lowest conduction band and $k_B$ is the Boltzmann constant. The $E_T$ changes from $\sim$\,1 meV to 36 meV as we go from the low T to the high T-range. This order of magnitude difference in $E_T$ suggests different capture mechanisms for the photo-generated carriers in different T-regions.

\subsection{Electrostatically tuned relaxation of the photoconducting state:}

We have discussed the electrostatic gate tuning of the dark conductivity of reduced STO in section-III(B) of this paper. Now, we will examine relaxation of the PC state under the gate field at 100 K and 300 K as shown in Fig. 7 and Fig. 8 respectively. At 100 K, first the R(t) profile under and after photo-illumination without any gate field was recorded. This was followed by four separate measurements of exposure and recovery in which a -\,3 kV/cm gate field was applied at t = 0, 100, 500 and 1000 seconds after turning off the photo-exposure. In each of these measurements, the sample was taken to room temperature after completion of each recovery cycle to ensure full recovery of the channel conductivity. In the 300 K measurement, the gate field applied at t = 0 (shown as $t_0$ in the figure) was switched off after relaxing the system slightly above the ground state. At this point ($t'_0$) the gate field was switched off and the system was allowed to come to its ground state value ($R_0$). Once the resistance saturates to $R_0$, the sample was subjected to photo-exposure again with the same photon flux. The $t_1$, $t_2$ and $t_3$ points shown in the Fig. 8(a) as dotted lines correspond to the time t = 100, 500 and 1000 seconds, respectively, when the gate field was applied. The dashed lines marked at $t'_1$, $t'_2$ and $t'_3$ in the figure mark the instant when the gate field was turned off. The results are presented in terms of normalized resistance, $R/R_0$; $R_0$ being the resistance in dark, as a function of time.

At 100 K (see Fig. 7), we observe an abrupt increase in R (jump at t = $t_n$, where n = 0 - 3) on application of negative $E_g$ and then a similar drop at $t'_n$ when the gate field is removed. However, this change in resistance is much smaller than the change caused by photo-illumination. On the other hand, the effect of negative $E_g$ on the recovery process at 300 K is remarkably different. Here the gate field not only increases the resistance abruptly at the point of application, but also accelerates the recovery process. Moreover, the identical slopes of the $R/R_0$ vs t curves in the time intervals ($t'_1 - t_1$), ($t'_2 - t_2$) and ($t'_3 - t_3$) suggest that the acceleration process is independent of the state of recovery. The process of recovery from the photoconducting state was also examined for several values of negative gate fields as shown in Fig. 8(b). Here the $E_g$ was applied at the same time when the photo-illumination was turned off. One can clearly see an increase in acceleration to recovery with the increasing field strength.
\begin{figure}[h]
\begin{center}
\includegraphics [ trim=0cm 0cm 0cm 0cm, width=8.7cm, angle=0 ]{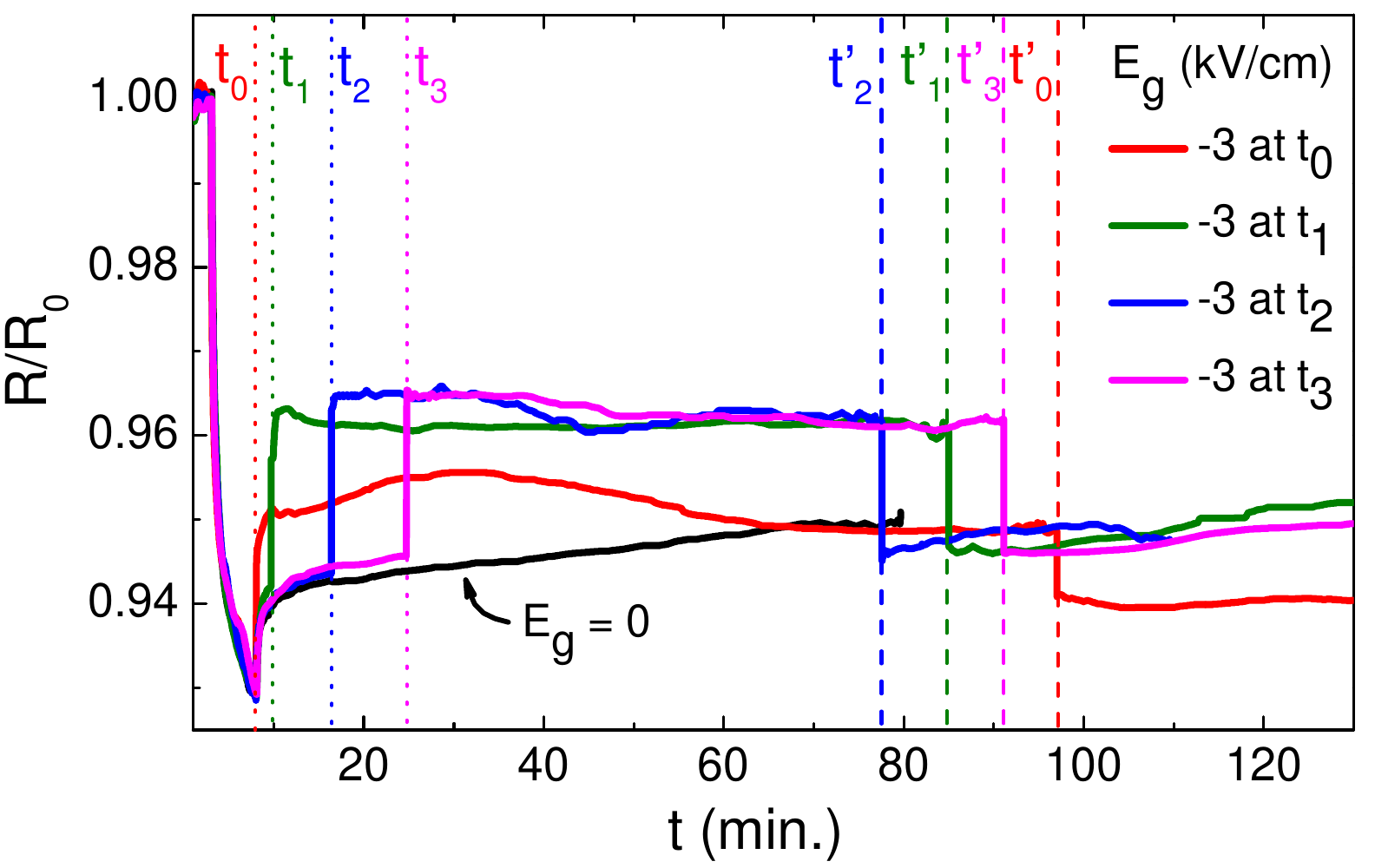}%
\end{center}
\caption{\label{fig 7} (Color online) The recovery processes under negative gate field at 100 K. Black solid line is the photo-response without any gate field and then the gate field of -\,3 kV/cm was applied during recovery at time t = 0 ($t_0$), 100 ($t_1$), 500 ($t_2$) and 1000 ($t_3$) seconds in four separate measurements, subsequently after turning off the light at $t_0$. The gate field applied at t = $t_0$, $t_1$, $t_2$ and $t_3$ was switched off at t = $t'_0$, $t'_1$, $t'_2$ and $t'_3$, respectively. Switching on the gate field (at t = $t_n$, where n = 0 to 3) leads to an abrupt increase in resistance while, there is a sudden drop in it when the gate field is turned off (at t = $t'_n$). The time scale of abrupt change in resistance is of the order of milliseconds.}
\end{figure}
\section{Discussion}

We first discuss the temperature dependence of the change in channel resistance on electrostatic gating and light exposure independently as shown in Fig. 3 and Fig. 4, respectively. The precipitous rise of $\Delta R_{E_g}$ below 100 K can be understood from the behaviour of the dielectric function $\epsilon$ of STO, also shown in Fig. 3. The back gate voltage (positive/negative) changes the number of carriers (add/withdraw) at the interface and hence affects the channel conductivity (increase/decrease). Also, the induced charge carriers modify the band near interface depending on the nature of the field as shown in the inset (b) of Fig. 3. This changes the confinement of charge carriers in the conducting layer. As an effect of negative gate field, a depletion layer is formed which reduces the effective spatial thickness of the conducting layer leading to a decrease in conductance. If the level of conduction band edge at the interface shifts by $\Delta E$ and $\lambda$ is the spatial extent over which the induced charge accumulates, then using Mott theory of field effect\cite{mott1979electronic}, the relative change in conductance for a semiconductor of thickness d can be written as,\\
\begin{equation}
\frac{\Delta G} {G_0} = \frac{\lambda}{d}\Big\{\frac{\alpha}{1+\alpha}F(-v_s)+\frac{1}{1+\alpha}F(v_s)\Big\}
\end{equation}
\begin{equation*}
F(v_s) = \sum_{m=1}^{\infty} \frac{(-v_s)^m} {mm!}
\end{equation*}
where, $\alpha$ is the ratio of electron to hole current and $v_s$ is the reduced surface potential $\Delta E/kT$. Also, for an exponentially decaying potential,
\begin{equation}
\lambda = \Big\{\frac{\epsilon} {4\pi e^2N(E_F)}\Big\}^{1/2}
\end{equation}
where, $N(E_F)$ is the density of states at the fermi level. From Eq. (1) and (2), one can infer a square root dependence of conductance on dielectric function. Since the dielectric function of STO increases dramatically below 100 K, one would expect $\Delta R_{E_g}$ to show a similar dependence.

We now discuss the temperature dependence of $\Delta R_{P}$. The gradual increase in PC on going down from 300 K to 100 K can be elucidated on the basis of $e^-$ - $h^+$ quasi-Fermi levels which move through the distributed levels leading to modification in the recombination kinetics\cite{sihvonen1967photoluminescence}, which increases the majority carrier life time on lowering the temperature. The precipitous increase in $\Delta R_P$ below 100 K must be linked to increase in mobility ($\mu$) or carrier density ($n_\Box$) or both. Many groups have reported enhanced $\mu$ on lowering the temperature to a value as high as $\approx$\,$10^4 cm^2/Vs$ at 10 K.\cite{yasunaga1968photo,Tufte1967electronmobility,Frederikse1964PhysRev,ishikawa2004isotope} Such increase in the mobility below 105 K may result from band widening due to increased overlap of 3d orbitals of Ti in the low temperature structural phase.\cite{katsu2000anomalous} However, no abrupt change in $\mu$ is seen at the AFD PT at 105 K.\cite{Tufte1967electronmobility} During this cubic to tetragonal PT, the tetragonal domains oriented along each of three original cubic axis result in the crystal with attendant twin boundaries. Recently, Kalisky et al. found the enhanced conductivity along these twin boundaries in LaAlO$_3$/SrTiO$_3$ heterointerfaces at 4 K.\cite{kalisky2013locally} It might be possible that the photo-generated carriers at the twin boundaries have greater mobility and hence it can contribute to the large photo-response seen at low temperatures in irradiated STO.
\begin{figure}[h]
    \centering
    %\subfigure[]{\includegraphics[height=10cm, width=18cm]{figure7.eps}}
    \subfigure{\includegraphics[ trim=0cm 0cm 0cm 0cm, width=8cm, angle=0 ]{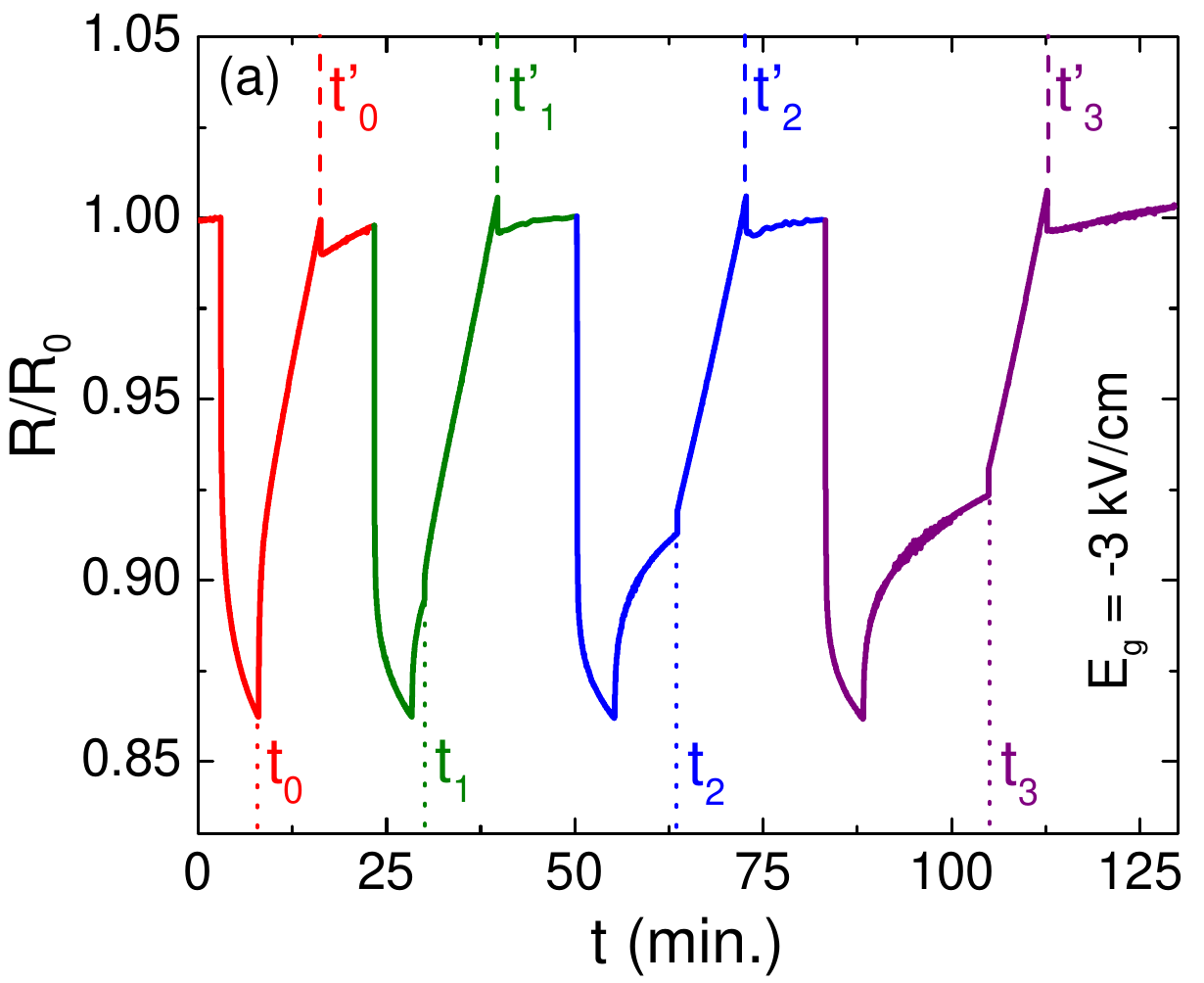}}
    \subfigure{\includegraphics[ trim=0cm 0cm 0cm 0cm, width=8.2cm, angle=0 ]{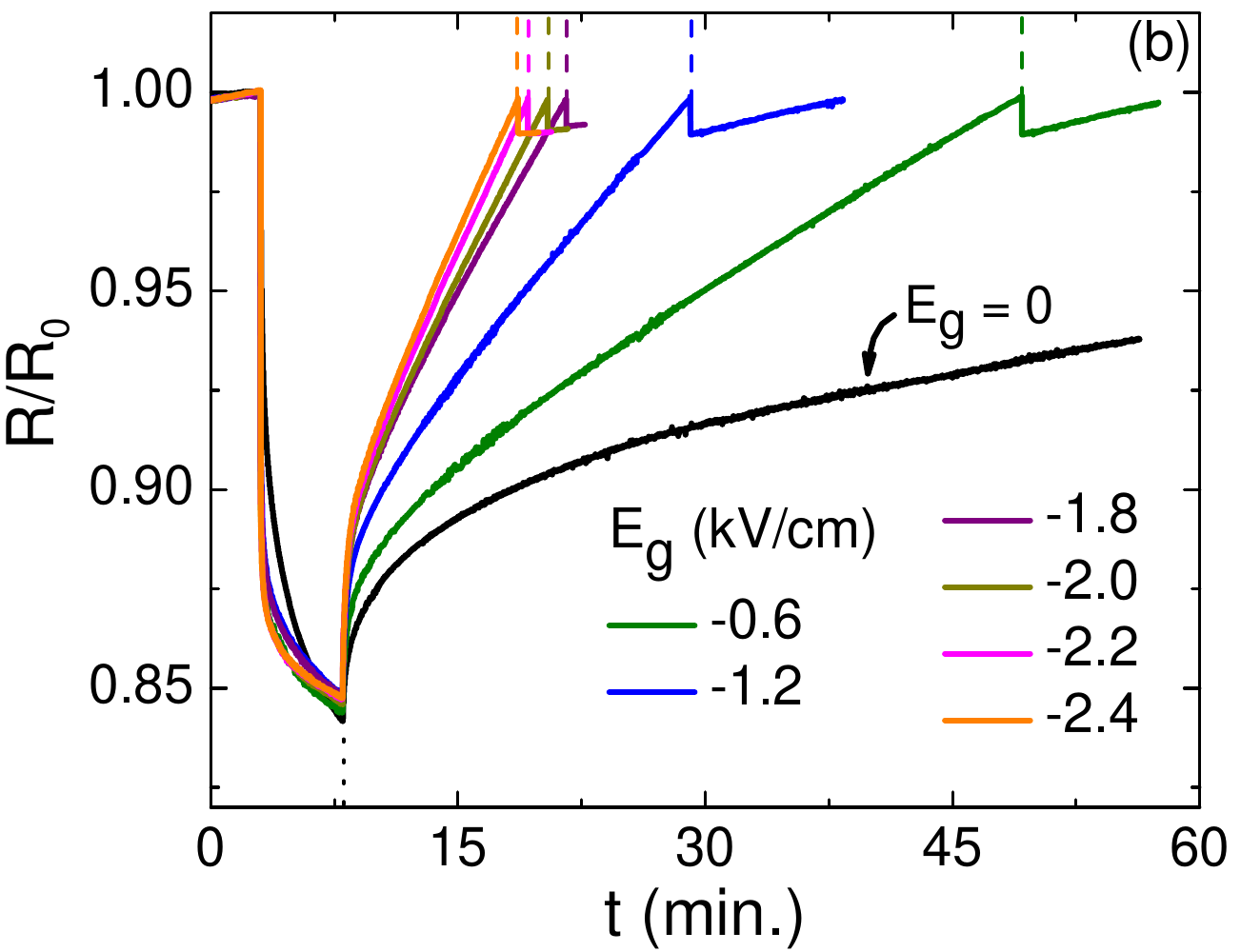}}
\caption{\label{fig 8} (Color online) The acceleration in recovery processes by negative gate field at 300 K. (a) The gate field of -\,3 kV/cm applied during recovery at time t = 0 ($t_0$) and left on till $t'_0$. The $R_\Box$ marginally exceeds its value in dark during $(t'_0 - t_0)$ seconds. The sample is allowed to recover its $R_\Box$ in dark after turning off the gate field and then reexposed under the same photon flux as earlier. This process is repeated for several values of $t_n$. (b) The dependence of recovery process on the strength of the negative gate field. The fields were applied at the time of switching off the light as marked by dotted line. The dashed lines indicate the point of turning off the field.}
\end{figure}

\begin{figure}[h]
\begin{center}
\includegraphics [ trim=0cm 0cm 0cm 0cm, width=8cm, angle=0 ]{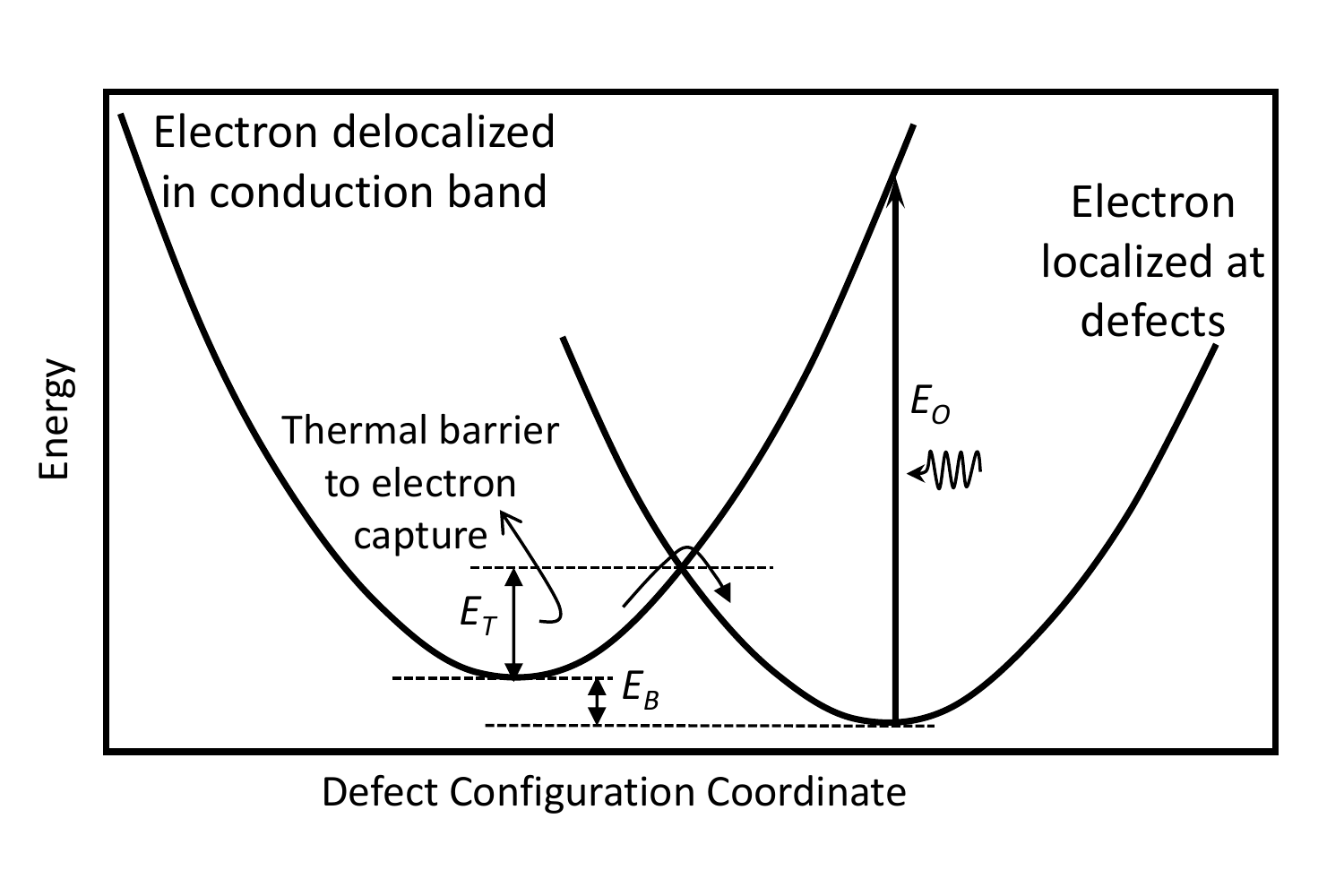}%
\end{center}
\caption{\label{fig 9} Defect configuration-coordinate diagram indicating the photo-excitation of an electron by energy $E_O$ from the defect into the conduction band, which is then captured in a thermal barrier of energy $E_T$. In order to relax in its ground state, the electron must have enough energy to surmount the barrier. The $E_B$ denotes the net binding energy gained in the process of large lattice relaxation.}
\end{figure}
Other likely mechanism for the enhanced PC is the higher photo absorption cross section in the AFD phase. The 105 K phase transition transforms the R $\rightarrow$ $\Gamma$ indirect optical transition in cubic phase to the much more effective direct $\Gamma$ $\rightarrow$ $\Gamma$ transition\cite{Mattheiss1972PRB,heifets2006calculations,Evarestov2011PRB,rossella2007photoconductivity}, which would lead to absorption of a good fraction of photons in the surface region of STO there by enhancing the sheet carrier density. It is also likely that the AFD-PT changes the nature of defect states in the gap which control carrier drift and recombination processes in the system.\cite{rossella2007photoconductivity,sihvonen1967photoluminescence}

At this stage, it is pertinent to discuss the persistent nature of photoconductivity and its behaviour under electrostatic gating. Theoretical approaches used to explain the phenomenon invoke either atomic scale microscopic barriers for recombination existing around defect centers with large lattice relaxation (LLR)\cite{Lang1977PRL}, or a spatial separation of photo-generated electrons and holes by macroscopic barriers due to band bending at the surface or interfaces.\cite{Queisser1979PRL} The ultraviolet photoemission spectroscopy studies on ion-bombarded STO show essentially zero band bending.\cite{Henrich1978PRB}
%While reduced STO is prone to chemisorption of oxygen,\cite{Henrich1978PRB,uedono2002study,Gentils2010PRB}
While reduced STO is prone to chemisorption of oxygen onto its surface which may lead to a non-zero band bending, the dominant role of thermal energy in PPC in high T-region along with a substantially week temperature dependence of the relaxation time in low T-region strongly suggest the existence of LLR dominated PPC. To discus this further, we invoke a defect configuration-coordinate diagram\cite{Lang1977PRL,Langandjaros} as sketched in Fig. 9, where a photon of energy $E_O$ promotes an electron from defect level to conduction band, which is re-captured by a thermal barrier preventing the recombination. Clearly, a photo-generated electron has to have enough thermal energy to cross this barrier for recapture onto the defect centers. The capture and thermal release model is consistent with our observation of a highly T-dependent behavior of decay time constant with an activation energy of $\sim$\,36 meV in the temperature range of 300 K to 100 K. However, below $\approx$\,100 K where the photo-carriers do not have enough thermal energy to surmount the barrier, the recapture may take place by multi-phonon emission (MPE) process.\cite{ridley1978multiphonon} In this case the capture rate becomes constant as T $\rightarrow$ 0. This is consistent with our observation of a nearly temperature independent $\tau$ below 100 K.
%(see 20 reference also from PRB lin)

We now attempt to understand how the PPC state which survives for several hours at ambient temperature is destroyed instantaneously by a -ve gate field. As the latter causes accumulation of holes at the interface of unirradiated (insulating) STO and the irradiated (conducting) STO, we believe these static charges lower the thermal barrier $E_T$ for electron capture as shown in Fig. 9 to a sufficient degree to facilitate thermally activated recombination at 300 K. This presumably ensures quick recovery to the state existing before illumination. Moreover, as clear from Fig. 8(b), the rate of recovery can be controlled by the field strength. These features can be useful in optoelectronic hybrid-memory devices, where optically we can achieve the process of writing information which can then be erased electrically in a controlled fashion. On the other hand, at 100 K, where the thermal energy is too low to release the trapped electrons sitting deep in the well, the negative gate field seems not to change the recovery process as such (see Fig. 7). The hole accumulation at the interface as an effect of negative gate field leads to an abrupt increase in resistance without affecting much the recovery process. Further studies are needed to understand the behaviour of recovery processes under the gate field.

\section{Conclusions}

In summary, the Ar$^+$ - ion irradiated STO (100) makes an interesting 2-dimensional metallic system displaying high carrier mobility at 15 K. We found evidence for the role of the anti-ferro distortive cubic-to-tetragonal phase transition in the temperature dependence of the photo-induced and electrostatic gating induced conductivity, which map with the temperature dependence of the STO dielectric function. A large UV sensitive persistent photoconductivity has been observed over the temperature range of 300 K to 15 K, with a relaxation time of several hours, similar to that seen in LAO-STO heterostructures. Interestingly, at 300 K, a negative gate field accelerates the post-illumination recovery process pushing the system to relax to its ground state quickly. This property has the potential for designing a solid state photoelectric switch. Our findings can be useful in understanding several unique properties of SrTiO$_3$ - other oxides (such as LaAlO$_3$) interfaces, where the oxygen vacancies can significantly affect the interfacial two-dimensional electron gas.

\begin{acknowledgements}
We thank A. Rastogi for helping in photoconductivity measurements and P. C. Joshi for technical help. D.K. acknowledges Indian Institute of Technology Kanpur and UGC-India for financial support. R.C.B. acknowledges J. C. Bose National Fellowship of the Department of Science and Technology, Government of India.
\end{acknowledgements}

\bibliography{references}

\end{document}